\documentclass[10pt,letterpaper]{article}         %
\usepackage{opex3}                                %
\usepackage{graphicx}
\usepackage{epsfig}
\usepackage{cite}
\usepackage{color}
\usepackage{amsmath,amssymb,subfigure}
\def\U#1{{\rm #1}} 
\def\u#1{_{\rm #1}}

\newcommand{\bra}[1]{\langle #1 |}
\newcommand{\ket}[1]{| #1 \rangle}

\def\H{{\rm H}}
\def\V{{\rm V}}

\def\00{\H\V}
\def\11{\V\H}

\begin{document}
\title{
Extracting an entangled photon pair from 
collectively decohered pairs at a 
telecommunication wavelength
}
\author{
Yoshiaki~Tsujimoto,$^{1,*}$
Yukihiro~Sugiura,$^{1}$
Makoto~Ando,$^{1}$\\
Daisuke~Katsuse,$^{1}$
Rikizo~Ikuta,$^{1}$
Takashi~Yamamoto,$^{1}$
Masato~Koashi,$^{2}$ and 
Nobuyuki~Imoto$^{1}$}

\address{
$^1$Graduate School of Engineering Science, Osaka University,
Toyonaka, Osaka 560-8531, Japan\\
$^2$Photon Science Center, 
The University of Tokyo, Bunkyo-ku, Tokyo 113-8656, Japan
}

\email{$^*$tsujimoto@mp.es.osaka-u.ac.jp} 



\begin{abstract}
We experimentally demonstrated entanglement extraction 
scheme by using photons at the telecommunication 
band for optical-fiber-based quantum communications. 
We generated two pairs of non-degenerate polarization 
entangled photons at 780~nm and 1551~nm by 
spontaneous parametric down-conversion and distributed 
the two photons at 1551~nm through a collective 
phase damping channel which gives the 
same amount of random phase shift on the two photons. 
Through local operation and classical communication, 
we extracted an entangled photon pair from two 
phase-disturbed photon pairs. An observed fidelity 
of the extracted photon pair to a maximally entangled 
photon pair was 0.73 $\pm$ 0.07 
which clearly shows the recovery of entanglement. 
\end{abstract}


\section{Introduction}
Faithful distribution of photonic entangled states 
between two distant parties is 
one of the significant issues in the field of quantum 
information processing such as quantum 
teleportation~\cite{PhysRevLett.70.1895}, 
superdense coding~\cite{PhysRevLett.69.2881} 
and entanglement-based quantum key 
distribution~\cite{PhysRevLett.68.557}. 
In order to perform these quantum information processing, 
schemes for robust distribution of entanglement are required. 
Entanglement distillation~\cite{PhysRevA.54.3824, 
PhysRevA.53.2046, PhysRevLett.77.2818} 
is one of the protocols for this purpose and several schemes 
have been proposed and demonstrated experimentally 
\cite{PhysRevA.64.012304, PhysRevA.64.014301, 
pan2001entanglement, PhysRevLett.90.207901, 
yamamoto2003experimental}. 
In previous proof-of-principle experiments, photons at 
visible wavelength were used. 
However, for practical optical-fiber-based quantum 
communication over a long distance, 
photons in telecommunication bands should be used to benefit 
from a low photon loss in an optical fiber. 

In this paper, we report an experimental demonstration 
of the entanglement extraction 
based on linear optics and a post-selection~\cite{PhysRevA.64.012304} 
by using two non-degenerated 
photon pairs at a visible wavelength of 780~nm and a telecom 
wavelength of 1551~nm. 
The sender Alice keeps the two visible photons and 
sends the two telecom photons to the receiver 
Bob through a collective phase damping channel~(CPC) 
which adds the same phase shifts to the two photons. 
Such a collective noise channel appears in many practical 
situations~\cite{stucki2002quantum, 
yamamoto2007experimental, PhysRevLett.92.257901} 
and discussed in many quantum communication protocols, 
such as reference-frame-free quantum communication~\cite{
PhysRevA.82.012304, wabnig2013demonstration, PhysRevLett.112.130501, 
PhysRevLett.113.060503}, decoherence-free quantum 
communication~\cite{kwiat2000experimental, PhysRevLett.92.107901, 
yamamoto2008robust, PhysRevLett.106.110503, PhysRevA.87.052325} 
and computation~\cite{PhysRevLett.91.217904, PhysRevLett.95.130501}. 
Through the transmission through the CPC, the entanglement of 
each photon pair is lost. 
However, after performing the quantum operation on two photons 
of 780~nm at Alice's side and a projective measurement on 
one of the two photons at Bob's side, 
Alice and Bob can extract an entangled photon pair. 
An observed fidelity of the extracted photon pairs to a maximally 
entangled state is 0.73 $\pm$ 0.07, which shows extraction of 
the entanglement from two phase-disturbed photon pairs shared between 
Alice and Bob. 

\section{The theory of the entanglement extraction}
We first introduce the demonstrated entanglement extraction scheme. 
\begin{figure}[t]
 \begin{center}
 \scalebox{0.5}{\includegraphics{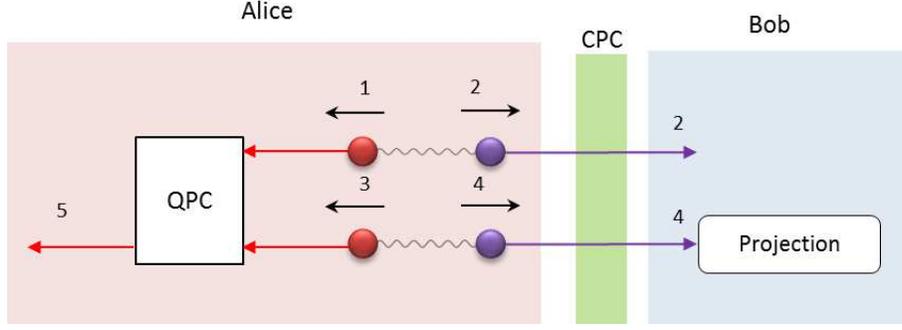}}
  \caption{(Color online)
  The schematic diagram of the entanglement extraction protocol. 
  Alice generates two polarization 
  entangled photon pairs and sends halves of the photon pairs to Bob
  through the collective phase-damping channel~(CPC). 
  When Alice performs the quantum parity check~(QPC) 
  on the two photons in modes 1 and 3, 
  and Bob performs a projection on the photon
  in mode 4, a maximally entangled photon pair is shared 
  in mode 2 and mode 5.
    \label{fig:scheme}}
 \end{center}
\end{figure}
The purpose of two parties, Alice and Bob, is to share 
a maximally entangled photon 
pair denoted by $\ket{\rm{\phi}^+}\equiv (\ket{\H\H}+\ket{\V\V})/\sqrt{2}$
, where $\ket{\H}$ and $\ket{\V}$ represent horizontal~($\H$) 
and vertical~($\V$) polarization states of a photon, respectively. 
As shown in Fig.~\ref{fig:scheme}, 
Alice generates $\hat{\rho}_{1234}=\hat{\phi}^{+}_{12}
\otimes\hat{\phi}^{+}_{34}$, where $\hat{\phi}^+\equiv\ket{\phi^+}\bra{\phi^+}$. 
The subscripts represent the spatial modes of the photons. 
She then sends the photons in modes 2 and 4 to Bob through a CPC. 
By denoting $\hat{Z}\equiv\ket{\H}\bra{\H}-\ket{\V}\bra{\V}$ 
and $\hat{Z}(\theta)\equiv\rm{exp}$$(-i\theta\hat{Z}/2)$, 
the CPC acting on the two photons 
in modes 2 and 4 transforms $\hat{\rho}_{1234}$ to 
\begin{eqnarray}
\hat{\rho}'\u{\rm{1234}}&=&\frac{1}{2\pi}\int 
d\theta\hat{Z}\u{2}(\theta)\otimes\hat{Z}\u{4}(\theta)\hat{\rho}_{1234}
\hat{Z}\u{2}^{\dagger}(\theta)\otimes\hat{Z}\u{4}^{\dagger}(\theta)\label{CPC}\\
&=&(\ket{\H\H\H\H}\bra{\H\H\H\H}+\ket{\H\H\V\V}\bra{\H\H\V\V}\nonumber
+\ket{\H\H\V\V}\bra{\V\V\H\H}+\ket{\V\V\H\H}\bra{\H\H\V\V}\nonumber\\
&+&\ket{\V\V\H\H}\bra{\V\V\H\H}+\ket{\V\V\V\V}\bra{\V\V\V\V})/4.
\end{eqnarray}
Because the density operator of each photon pair 
$\rm{Tr}\u{34(12)}[\hat{\rho}'\u{1234}]
=(\ket{\U{HH}}\u{34(12)}\bra{\U{HH}}\u{34(12)}
+\ket{\U{VV}}\u{34(12)}\bra{\U{VV}}\u{34(12)})/2$ 
has no entanglement, Alice and Bob do not share 
any entangled photon pairs as long as 
they treat the photon pairs separately. 
However, they can extract a maximally entangled state 
of a photon pair from $\hat{\rho}'\u{1234}$ of the whole 
system by local operation and classical 
communication in the following way. 

Alice performs the quantum parity check (QPC)~\cite{PhysRevA.64.062311} 
on photons in modes 1 and 3. 
Kraus operators of the QPC in Fig.~\ref{fig:scheme} 
is described as $\{\hat{F},\sqrt{\hat{I}-\hat{F}^\dagger\hat{F}} \}$, 
where $\hat{F}\equiv\ket{\U{H}}\u{5}\bra{\U{HV}}\u{13}
+\ket{\U{V}}\u{5}\bra{\U{VH}}\u{13}$ corresponds to a successful 
operation of the QPC and the other Kraus operator to a failure operation. 
After the successful operation of the QPC, the quantum state becomes 
$\hat{\rho}\u{\rm{QPC}}\equiv\hat{F}\hat{\rho}'\u{1234}\hat{F}^{\dagger}/\rm{Tr}$
$[\hat{F}^{\dagger}\hat{F}\hat{\rho}'\u{1234}]
=(\ket{\H \H \V}\u{524}\bra{\H \H \V}\u{524}
+\ket{\V \V \H}\u{524}\bra{\H \H \V}\u{524}+
\ket{\H \H \V}\u{524}\bra{\V \V \H}\u{524}
+\ket{\V \V \H}\u{524}\bra{\V \V \H}\u{524}$)/2. 
When Bob performs a projective measurement 
$\{\ket{+}\bra{+}, \ket{-}\bra{-}\}$ on the photon 
in mode 4, where $\ket{\pm}\equiv(\ket{\H}\pm\ket{\V})/\sqrt{2}$, 
the state in modes 2 and 5 is projected onto $\ket{\phi^{+}}\u{25}$ 
or $\ket{\phi^{-}}\u{25}
\equiv(\ket{\H\H}-\ket{\V\V})/\sqrt{2}$. 
By performing a phase flip operation when the photon in mode 
4 is projected onto $\ket{-}$, 
$\ket{\phi^-}$ is corrected to $\ket{\phi^+}$. 
As a result, Alice and Bob share $\ket{\phi^+}$ 
with a success probability $\rm{Tr}$
$[\hat{F}^\dagger\hat{F}\hat{\rho}\u{\rm{QPC}}]$=1/4. 
In our experiment, we do not perform the phase flip 
operation, and the success probability becomes 1/8. 
We notice that in the scheme, the effect of the 
phase disturbance on the photons in modes 2 and 4 
is compensated by the QPC on the photons 1 and 3 
at Alice's side. 
Such a non-local cancellation of the phase 
fluctuation is the result of the use of a significant 
property of the maximally entangled states, 
which is described by the relation 
$(\hat{I}\u{1(3)}\otimes \hat{Z}\u{2(4)})\ket{\phi^+}\u{12(34)}
=(\hat{Z}\u{1(3)}
\otimes \hat{I}\u{2(4)})\ket{\phi^+}\u{12(34)}$~\cite{PhysRevLett.106.110503, PhysRevA.87.052325}. 

\section{Experiment}
\subsection{Experimental setup}
\begin{figure}[t]
 \begin{center}
 \scalebox{0.37}{\includegraphics{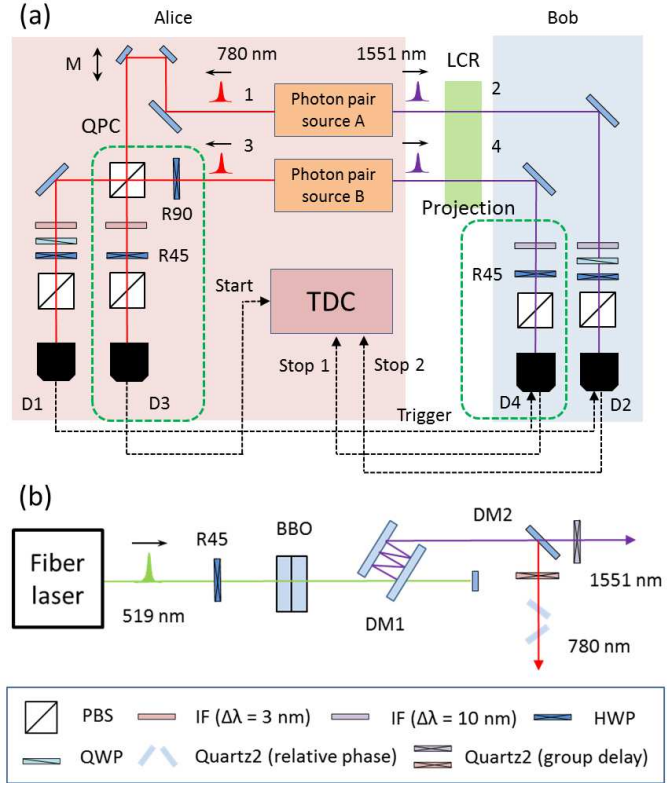}}
  \caption{(Color online) 
  (a)~The experimental setup for the entanglement extraction. 
  The half wave plate~R90 transforms
  $\ket{\H}$ to $\ket{\V}$ and vice versa. 
  The half wave plate~R45 transforms $\ket{\H}$ to $\ket{+}$ and 
  $\ket{\V}$ to $\ket{-}$, and vice versa. 
  (b)~The experimental setup of photon sources A or B. 
  The pulsed pump light~(519~nm) is 
  obtained by frequency doubling the output light of 
  the mode locked fiber laser at 1037~nm. 
  The details are shown in the main text. 
        \label{fig:EPPsetup}}
 \end{center}
\end{figure}
The experimental setup is shown in Fig.~\ref{fig:EPPsetup}~(a). 
Two entangled photon pairs 
are generated by spontaneous parametric down conversion~(SPDC) 
at source A and B. 
The setup of the photon pair sources is shown in 
Fig.~\ref{fig:EPPsetup}~(b). 
The pump laser is based on a fiber laser~(wavelength: 
1037~nm, pulse width: 381~fs, 
repetition rate: 80~MHz), which is frequency doubled 
such that the center wavelength is 
519~nm with the power of 1.32~W. 
The polarization of the pump beam is set to be diagonal 
to the axes of two adjacent 
phase-matched 1-mm-thick $\beta$-barium borate~(BBO) 
crystals by the half wave plate~(R45). 
In this experiment, we select the non-degenerate photon 
pair at a visible wavelength of 780~nm 
and a telecom wavelength of 1551~nm among the photon pairs 
generated from the BBOs. 
After the BBO crystal, the pump beam is removed by using a 
pair of dichroic mirrors~(DM1) 
whose reflectance for the photons at 780~nm and 1551~nm 
is over $99~\%$ 
and transmittance for the photons at 519~nm is $\sim97~\%$. 
After DM1, the visible photons 
and telecom photons are separated into different spatial modes by DM2. 
For both photons, the group delays between $\ket{\H}$ and 
$\ket{\V}$ are compensated by 
6.17~mm and 7.44~mm-thick quartz crystals~(Quartz1)~
in the paths of the visible and the telecom photons, 
respectively. 
The relative phase shift between $\ket{\H\H}$ and $\ket{\V\V}$ 
is compensated by 
tilting 0.6~mm-thick quartz crystals~(Quartz2). 

As shown in Fig.~\ref{fig:EPPsetup}~(a), 
two visible photons from photon source A and B 
are injected into a polarizing beamsplitter~(PBS) 
simultaneously by adjusting a moving mirror~(M). 
The spectra of the two output photons from the PBS are 
filtered by interference filters~(IFs) 
with a bandwidth of 3 nm, and then coupled to single-mode 
fibers followed by silicon avalanche 
photodetectors~(quantum efficiency: 60 $\%$) D1 and D3. 
On the other hand, two telecom photons pass through a 
liquid crystal retarder~(LCR). 
The LCR adds eight phase shifts $n\pi/4~(n=0, 1, 2 , 
\cdots, 7)$ between $\ket{\H}$ and $\ket{\V}$ 
of the two photons by switching the applied voltage. 
The operations of the LCR on the two photons 
are nominally described by the Kraus operators 
$\{\hat{Z}^{n \pi/4}_2 \otimes \hat{Z}^{n \pi/4}_4\}_{n=0, 1, 2 , \cdots, 7}$. 
Because this transformation is exactly the same 
completely positive and trace preserving 
map as the one given in Eq.~(\ref{CPC}), 
the LCR simulates the CPC. 
The telecom photons are sent through the interference 
filters~(IFs) with a bandwidth of 10~nm and 
then they are coupled to single-mode fibers followed by 
two InGaAs avalanche photodetectors~(quantum 
efficiency: 25 $\%$) D2 and D4 which are gated by 
the electric signal from D1. The electric 
signal from D3 is connected to a time-to-digital 
converter~(TDC)~as a start signal, and electric 
signals from D2 and D4 are used as stop signals of 
the TDC. 
We postselect the events where two stop signals are 
clicked within the time difference of 2~ns, 
which guarantees the four-fold coincidence among 
D1, D2, D3 and D4.

\subsection{Experimental results}
\begin{figure}[t]
 \begin{center}
 \scalebox{0.4}{\includegraphics{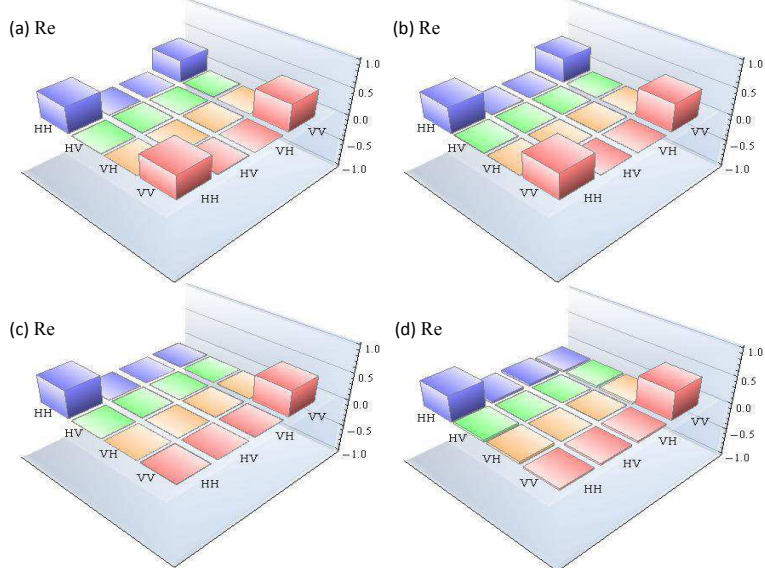}}
  \caption{(Color online) 
  The real parts of the matrix elements of 
  (a)~$\hat{\rho}\u{12}$, (b)~$\hat{\rho}\u{34}$, 
  (c)~$\hat{\rho}'\u{12}$ and (d)~$\hat{\rho}'\u{34}$.
    \label{fig:matrix}}
 \end{center}
\end{figure}

We first characterized the initial two photon pairs from 
photon pair sources A~($\hat{\rho}\u{12}$) 
and B~($\hat{\rho}\u{34}$). 
By performing the quantum state tomography~\cite{PhysRevA.64.052312} 
with diluted maximum-likelihood 
algorithm~\cite{PhysRevA.75.042108}, we reconstructed the density 
operators of the photon pairs as shown 
in Figs.~\ref{fig:matrix}~(a) and (b). 
Observed fidelities of $\hat{\rho}\u{12}$ and 
$\hat{\rho}\u{34}$ to $\ket{\phi^+}$ were 0.92 
$\pm$ 0.01 and 0.94 $\pm$ 0.01, respectively, 
which implies that the two photon pairs were highly entangled. 
The detection rates of $\hat{\rho}\u{12}$ and 
$\hat{\rho}\u{34}$ were 920 Hz and 620 Hz, respectively. 
We estimated that the photon pair generation rate is of 
the order of $10^{-2}$, which indicates 
that the multi-photon emission from the photon 
sources is negligibly small.

Next we reconstructed the density operators when 
we applied phase fluctuations to each photon pair 
by using the LCR. 
Figures.~\ref{fig:matrix}~(c) and (d) show the reconstructed 
density matrices of $\hat{\rho}'\u{12}$ 
and $\hat{\rho}'\u{34}$ with phase fluctuations. 
Off-diagonal elements of the density matrices disappeared, 
which indicates that the LCR effectively simulates 
the phase-damping channel. Observed fidelities of 
$\hat{\rho}\u{12}$ and $\hat{\rho}\u{34}$ to 
$\ket{\phi^+}$ were 0.50 $\pm$ 0.01 and 0.46 
$\pm$ 0.02, respectively.

\begin{figure}[t]
 \begin{center}
 \scalebox{0.78}{\includegraphics{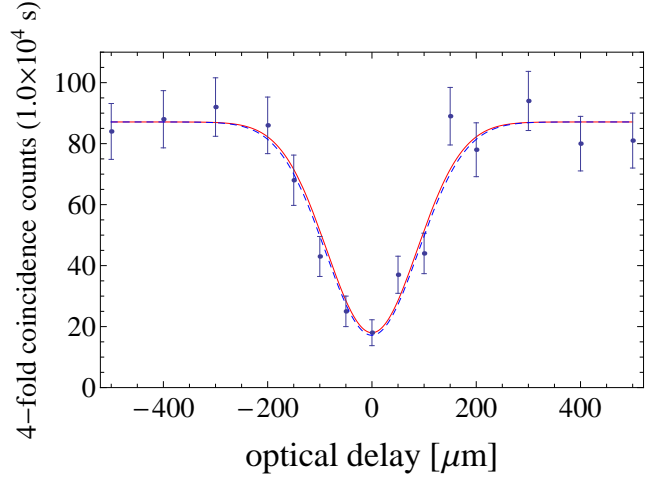}}
  \caption{(Color online) 
  The observed Hong-Ou-Mandel interference between 
  two visible photons in modes 1 and 3. 
  Each point was recorded for $1.0\times10^4$ s. 
  The red solid curve is the Gaussian fit to the 
  obtained data. The blue dashed curve is 
  obtained by Eq.~(\ref{theory}) with experimental 
  parameters. 
  \label{fig:HOMdip}}
 \end{center}
\end{figure}

Before demonstrating the entanglement extraction 
scheme, we performed the Hong-Ou-Mandel~(HOM) 
interference~\cite{PhysRevLett.59.2044} of the photons in modes 
1 and 3 which are heralded by the photon 
detection at D4 and D2, respectively. 
The experimental result is shown in 
Fig.~\ref{fig:HOMdip}. 
We clearly observed the HOM dip around zero 
delay point. 
The observed visibility at zero delay was 0.80 
$\pm$ 0.05. 
The full width at the half maximum~(FWHM) was 
calculated as $\sim$ 204 $\rm{\mu m}$ by fitting 
the experimental data with the Gaussian. 

\begin{figure}[t]
 \begin{center}
 \scalebox{0.35}{\includegraphics{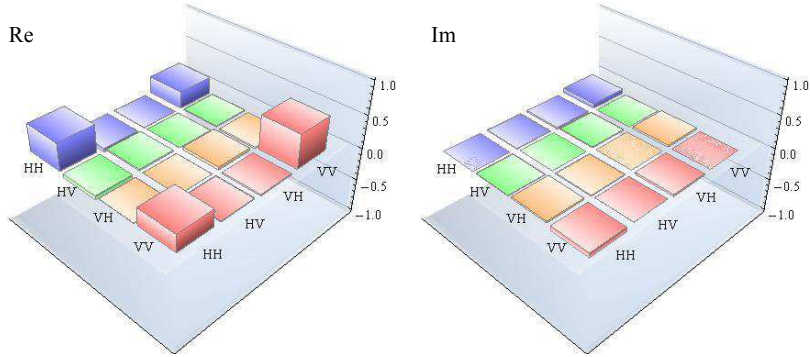}}
  \caption{(Color online) 
  (left)~the real part and (right)~imaginary 
  part of the density operator 
  $\hat{\rho}\u{\rm{final}}$ of the extracted 
  photon pair.
    \label{fig:matrix2}}
 \end{center}
\end{figure}

Finally, we reconstructed the density operator 
$\hat{\rho}\u{\rm{final}}$ when we performed 
the entanglement extraction scheme on the two photon 
pairs. 
The experiment was performed at the zero delay point 
of the HOM dip in Fig.~\ref{fig:HOMdip}.
The reconstructed density matrix is shown in 
Fig.~\ref{fig:matrix2}. 
An observed fidelity of $\hat{\rho}\u{\rm{final}}$ 
to $\ket{\phi^+}$ was 0.73 $\pm$ 0.07, 
clearly exceeding the threshold value of 0.5 to show 
that the extracted pair of photons are entangled. 
This means that entanglement was successfully 
distributed at the telecom regime in the presence 
of corrective phase fluctuations. 

\section{Discussion}
In the following, we consider the reason for the 
degradation of the observed values of the 
visibility of the HOM interference between 
the photons in modes 1 and 3 and the fidelity 
of $\hat{\rho}\u{\rm{final}}$ after the 
entanglement extraction scheme. 
Because the probability of the multi-photon 
pair emission is sufficiently low, 
the two-photon state obtained by SPDC after 
IFs is expressed as 
\begin{equation}
\ket{\Psi}_{ij}\simeq \iint d\omega d\omega'
\Phi(\omega,\omega')\hat{a}^{\dagger}_{i}
(\omega)\hat{a}^{\dagger}_{j}(\omega')\ket{\mathrm{vac}},
\end{equation}
where 
$\ket{\rm{vac}}$ is the vacuum state, and 
$\hat{a}_{k}^\dagger(\omega)$ is a creation operator 
at the angular frequency of 
$\omega$ in spatial mode $k(=1,2,3,4)$. 
$\Phi(\omega,\omega')$ is a product of the spectral 
amplitude of the two-photon state generated from the SPDC and
the transmission coefficients of the IFs for the visible 
and telecom photons. 
Because the detectors distinguish neither angular 
frequencies nor exact arriving times of the photons, 
the four-fold coincidence probability $P_{1234}$ 
among the detectors D1, D2, D3 
and D4 is given by the sum of all the frequency 
contributions, 
which is proportional to $\iiiint d\omega\u{1}d\omega\u{2}
d\omega\u{3}d\omega\u{4}|\bra{\mathrm{vac}}
\hat{a}\u{1}(\omega\u{1})\hat{a}\u{3}(\omega\u{3})
\hat{U}\ket{\phi(\omega\u{2})}\u{1}\ket{\phi(\omega\u{4})}\u{3}|^2.$ 
Here 
$\ket{\phi(\omega\u{2})}\u{1}\equiv\bra{\mathrm{vac}}
\hat{a}\u{2}(\omega\u{2})\ket{\Psi}\u{12}
=\int d\omega\Phi(\omega,\omega\u{2})\hat{a}^{\dagger}
\u{1}(\omega)\ket{\mathrm{vac}}$ 
and $\ket{\phi(\omega\u{4})}\u{3}\equiv\bra{\mathrm{vac}}
\hat{a}\u{4}(\omega\u{4})\ket{\Psi}\u{34}
=\int d\omega\Phi(\omega,\omega\u{4})\hat{a}^{\dagger}\u{3}(\omega)
e^{i\omega\tau}\ket{\mathrm{vac}}$, 
where $\tau$ is the time delay in mode~3. 
The unitary operator $\hat{U}$ represents the transformation 
at the HBS, which satisfies
$\hat{U}\ket{\mathrm{vac}}=\ket{\mathrm{vac}}$, 
$\hat{U}\hat{a}^{\dagger}\u{1}(\omega)\hat{U}^{\dagger}
=(\hat{a}^{\dagger}\u{1}(\omega)+\hat{a}^{\dagger}\u{3}(\omega))/\sqrt{2}$ and 
$\hat{U}\hat{a}^{\dagger}\u{3}(\omega)\hat{U}^{\dagger}
=(\hat{a}^{\dagger}\u{1}(\omega)-\hat{a}^{\dagger}\u{3}(\omega))/\sqrt{2}$. 
In our experiment, we assume that phase matching 
bandwidth of the BBO crystal is sufficiently broad, 
and the spectral distribution function of the pump 
beam for the photon pairs is 
a Gaussian with a variance $\delta\omega_{p}$ and 
a center angular frequency $\omega_{p}$, and 
those of the IFs for the visible/telecom photons 
are Gaussians with a 
variance $\delta\omega_{v/t}$ and a center angular 
frequency $\omega_{v/t}$. 
By these assumptions, $\Phi(\omega,\omega')$ is represented as 
$\Phi(\omega,\omega')=\rm{exp}
[-(\omega+\omega'-\omega_p)^2/(4\delta\omega^2_p)]
\rm{exp}
[-(\omega-\omega_{v})^2/(4\delta\omega^2_{v})]
\rm{exp}
[-(\omega'-\omega_{t})^2/(4\delta\omega^2_{t})]$. 
Then $P_{1234}$ is calculated as 
\begin{eqnarray}
P\u{1234} \propto 1 - \sqrt{\frac{\delta\omega\u{p}^2
(\delta\omega\u{p}^2+\delta\omega\u{v}^2
+\delta\omega\u{t}^2)}{(\delta\omega\u{v}^2+\delta\omega\u{p}^2)
(\delta\omega\u{t}^2+\delta\omega\u{p}^2)}}
e^{\frac{-\delta\omega^2\u{v}\delta\omega^2\u{p}\tau^2}
{\delta\omega^2\u{v}+\delta\omega^2\u{p}}}.
\label{theory}
\end{eqnarray} 
The coefficient of the second term in right hand side of 
Eq.~(\ref{theory}) is the visibility of the HOM dip. 
In our experiment, we use the pump beam at 519~nm with a pulse 
width of 397~fs, the interference filters with a bandwidth of 
3~nm for the visible photons at 780~nm 
and those with a bandwidth of 10~nm for the telecom photons 
at 1551~nm. 
These values correspond to $\delta\omega_p\simeq 3.0\times10^{12}~\rm{s}^{-1}$, 
$\delta\omega_v\simeq 3.9\times10^{12}~\rm{s}^{-1}$ and $\delta\omega_t\simeq 
3.3\times10^{12}~\rm{s}^{-1}$. 
The dependency of the visibility 
on $\tau$ is predicted as in Fig.~\ref{fig:HOMdip}. 
The visibility at zero delay and the FWHM of the dip are 
found to be 0.80 and 210~$\rm{\mu m}$. 
We see that the theoretical curve is in good agreement 
with the experimental results. 
Thus we expect that a higher visibility will be obtained 
by using a narrower spectral filtering.

Next, we consider the reason for the degradation of 
the fidelity of $\hat{\rho}\u{\rm{final}}$ after 
the entanglement extraction scheme.
We assume that the visibility of the HOM interference 
does not depend on the polarization of the photons. 
We also assume that 
each input pulse is described by a photon in a single 
temporal and spatial mode, 
and the overlap~(indistinguishability) between the 
two  mode shapes is given 
by the observed visibility of 0.80. 
By using the reconstructed density operators of 
$\hat{\rho}\u{12}$ and $\hat{\rho}\u{34}$ 
in Figs.~\ref{fig:matrix}~(a) and~(b) as the 
initial photon pairs, 
the fidelity of the two-photon state after the 
entanglement extraction is calculated to be 0.79, 
which is within the margin of the statistical 
error of our experimental result. 
In this model, the degradation of the fidelity 
is caused by the imperfection of the initial 
state and the visibility of the HOM interference. 
If we prepare the maximally entangled states 
$\hat{\phi}^+\u{12}$ and $\hat{\phi}^+\u{34}$ as 
the initial photon pairs and the visibility of HOM 
interference is 0.80, 
the fidelity of the extracted state is calculated 
to be 0.90. If we use the narrower IFs such 
that the visibility is 1, 
the fidelity of the final two-photon state is 
calculated to be 0.87 from the initial states 
$\hat{\rho}\u{12}$ and $\hat{\rho}\u{34}$. 

\section{Conclusion}
In conclusion, we have demonstrated the entanglement 
extraction scheme using the polarization 
entangled photon pairs at the visible wavelength of 
780~nm and the telecom wavelength of~1551 nm. 
The observed fidelity of the extracted photon pair 
is 0.73 $\pm$ 0.07, which clearly shows 
the recovery of the entanglement shared between the 
parties. While this scheme was demonstrated for 
distributing entangled photon pairs against collective 
phase noise, 
it is also applicable to the entanglement distribution 
against 
general collective noises for two qubits by sending 
H-polarized and V-polarized photons through different 
channels as proposed in Ref.~\cite{yamamoto2007experimental}.
We believe that our result is useful for the efficient and 
robust distribution of entanglement 
through optical fibers over a long distance. 

\section*{Acknowledgments}
This work was supported by the Funding Program for World-Leading 
Innovative R\&D on Science and Technology (FIRST), MEXT 
Grant-in-Aid for Scientific Research on Innovative Areas 21102008, 
JSPS Grant-in-Aid for Scientific Research(A) 25247068, 
(B) 25286077 and (B) 26286068.
\end{document}